\documentclass{osa-article}

\journal{osajournal} 
\setprjcopyright
\articletype{Research Article}

\begin{document}

\title{Femtosecond nonlinear losses in multimode optical fibers}

\author{Mario Ferraro,\authormark{1,*} Fabio Mangini,\authormark{2} Mario Zitelli,\authormark{1} Alessandro Tonello,\authormark{3} Antonio De Luca,\authormark{4,5} Vincent Couderc,\authormark{3} and Stefan Wabnitz\authormark{1,6}}

\address{\authormark{1}Department of Information Engineering, Electronics, and Telecommunications, Sapienza University of Rome, Via Eudossiana 18, 00184 Rome, Italy\\
\authormark{2}Department of Information Engineering, University of Brescia, Via Branze 38, 25123 Brescia, Italy\\
\authormark{3}Université de Limoges, XLIM, UMR CNRS 7252, 123 Avenue A. Thomas, 87060 Limoges, France\\
\authormark{4}Physics Department, University of Calabria, I-87036 Arcavacata di Rende, CS, Italy\\
\authormark{5}CNR Nanotec-Institute of Nanotechnology, S.S. Cosenza, I-87036 Rende, CS, Italy\\
\authormark{6}CNR-INO, Istituto Nazionale di Ottica, Via Campi Flegrei 34, I-80078 Pozzuoli (NA), Italy\\}

\email{\authormark{*}mario.ferraro@uniroma1.it} 

\begin{abstract}
Research on multimode optical fibers is arousing a growing interest, for their capability to transport high-power laser beams, coupled with novel nonlinear optics-based applications. However, when beam intensities exceed a certain critical value, optical fiber breakdown associated with irreversible modifications of their refractive index occurs, triggered by multiphoton absorption. These processes have been largely exploited for fiber material microstructuration. Here we show that, for intensities slightly below the breakdown threshold, nonlinear absorption strongly affects the dynamics of a propagating beam as well. We experimentally analyze this sub-threshold regime, and highlight the key role played by spatial self-imaging in graded-index fibers for enhancing nonlinear optical losses. We characterize the nonlinear power transmission properties of multimode fibers for femtosecond pulses propagating in the near-infrared spectral range. We show that an effective N-photon absorption analytical model is able to describe well the experimental data. 
\end{abstract}

\section{Introduction}
Nonlinear optics in multimode optical fibers (MMFs) is an emerging research field, for it leads to new possibilities for the control of the spatial, temporal and spectral properties of ultrashort light pulses~\cite{krupa2019multimode}.
Differently from the case of single-mode fibers, which are limited by their small transverse core size, large area MMFs permits for scaling up by orders of magnitude their energy transport capabilities. 
As a result, research in MMFs has aroused a growing interest for a variety of technologies, e.g., high-power fiber lasers~\cite{krupa2019multimode}, supercontinuum light sources~\cite{lopez2016visible}, high-resolution biomedical imaging~\cite{moussa2020spatiotemporal}, and micromachining~\cite{norman2006power}. From a fundamental viewpoint, the high beam intensity that can be reached in MMFs has also led to unveiling different complex nonlinear phenomena~\cite{Picozzi2015R30, mangini2020giving}.

The optical power transmission of MMFs is limited by different nonlinear effects, whose relative relevance strongly depends on the time scale of the propagating pulses.
As well known, in the continuous-wave (cw) (or quasi-cw) regime the power transmission limitation of optical fibers is set by Brillouin scattering~\cite{Agrawal01}. When sub-nanosecond are employed, different nonlinear loss effects may become relevant.
Pulses longer than tens of picoseconds promote valence electrons to the conduction band: electron-phonon interactions result in a heating of the fiber material. Injecting intense light pulses in optical fibers leads to thermally-induced irreversible damages, such as fiber melting and boiling~\cite{stuart1995laser,stuart1996nanosecond}. Whenever the temperature reaches extreme values, typically above thousands of Kelvin, the fiber fuse phenomenon takes place \cite{shuto2004fiber}.
On the other hand, when intense femtosecond pulses are injected into MMFs, the fiber may break down due to ionization phenomena: these typically occur when the laser intensity is above $10^{12} W/cm^2$~\cite{cho1998observation}. Under proper conditions, photoionization leads to material modifications and ablation. These phenomena have been widely exploited over the past decades, mainly for their application to glass micromachining, e.g., for fabricating fiber Bragg gratings~\cite{mihailov2011bragg}.

Owing to the presence of a bandgap, ionization mechanisms in dielectric materials are triggered by multiphoton absorption (MPA) processes ~\cite{lenzner1998femtosecond, wu2005femtosecond}.
MPA plays a significant role even below the breakdown threshold, and it represents a major drawback for optical beam delivery, for it limits the efficiency of optical elements in the high-power regime. In fused silica, which is the main constituent of commercial MMFs, MPA may lead to nonlinear contributions to both the refractive index and the absorption coefficient, when operating at wavelengths between the visible and the mid-infrared range~\cite{sheik1990sensitive,boudebs2009absolute}. Because of the high intensities associated with the occurrence of MPA, its presence is generally neglected in a telecom context. Indeed, many photons need to be simultaneously absorbed to fill the pure silica bandgap (> 10 eV) at telecom frequencies: up to six-photons absorption has been observed at $\lambda = 790$  nm~\cite{cho2002situ}.

However, the presence of material defects, whose absorption band peaks at a few electronvolts, lowers the number of photons needed for the observation of MPA. For undoped core silica MMFs, the dominant defects, i.e., those with the highest oscillation strength, are the so-called Non-Bridging Oxygen Hole Centers (NBOHC) \cite{girard2019overview}. Recently, it has been shown that pulses at $\lambda =$ 1030 nm leads to the simultaneous absorption of 5-photons by NBOHCs. This is followed by the generation of a visible photoluminescence (PL) at the self-focusing point, where the intensity reaches its peak values~\cite{ManginiFerraro2020PRApp}. 

In the particular case of graded-index fibers (GIFs), the Germanium doping used to shape the core refractive index widens the range of defects, whose oscillation strength is comparable with that of NBOHCs. These defects are the Ge-related Oxygen Deficiency Centers (Ge-ODC), whose main characteristic is their blue-violet luminescence, and the paramagnetic Ge(1) and Ge(2) centers, that do not show any PL emission \cite{girard2019overview}. The PL of GIF defects permits to directly visualize the spatial self-imaging (SSI) phenomenon, the latter being the periodic replication of the electromagnetic field upon its propagation ~\cite{allison1994observations, singolomodo}. SSI is particularly studied in GIFs, since its period is remarkably short (a few hundreds microns), owing to the parabolic refractive index shape, which leads to equally spaced propagation constants for nondegenerate modes~\cite{mafi2012pulse}. As a result, a laser beam propagating inside a GIF continuously widens and tightens its waist: correspondingly PL generates a periodic array of light spots \cite{hansson2020nonlinear}.

In this work, we investigate ultrashort pulse propagation in MMFs, in a power regime which is close but still below their breakdown value. Here MPA mechanisms turn out to be relevant, but no damages of the fiber material are induced yet. This permits to carry out a detailed experimental characterization of optical nonlinear losses (NL), which occur both in the normal and in the anomalous dispersion regime of propagation. We first verify that thermal effects are not involved in determining such losses. Next, by comparing NL measurements of GIFs and step-index fibers (SIF), we point out that SSI has a key role in enhancing those losses. We investigate the origin of NLs, by comparing the wavelength dependence of the nonlinear fiber transmission properties with that of defects PL intensity. More specifically, we study the fiber power transmission properties when varying the fiber length, as well as the input beam wavelength and pulsewidth. We found that NL cannot be fully ascribed to the MPA which is responsible for PL. To reproduce our observations, we introduce an analytical model, where an effective N-photon absorption term is added to the propagation equation. A value of $N\simeq3$ is obtained, in good agreement with the experimental transmission at $\lambda =$ 1030 nm. Our results are important, as the may provide a guideline for modeling the power scaling of spatiotemporal ultrashort pulse propagation in MMFs. 


\section{Experiments}
\begin{figure}[!h]
\centering\includegraphics[width=8.4cm]{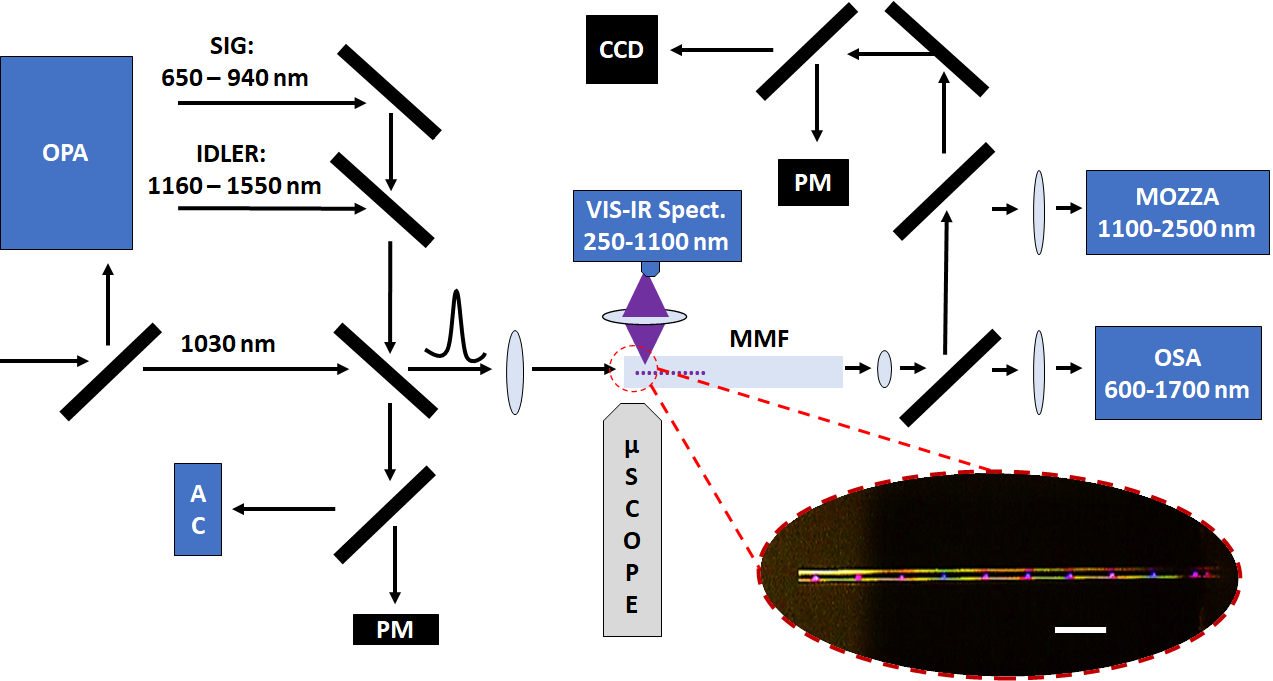}
\caption{Experimental set-up to characterize the NL of MMFs. In the inset, we show a microscope image of the scattered PL in correspondence of the self-imaging points of a 50/125 GIF. The white scale bar is 500 $\mu$m long.}
\label{set-up}
\end{figure}

The experimental setup to study NL phenomena in MMFs is shown in Fig.\ref{set-up}. It consists of an ultra-short laser system, involving a hybrid optical parametric amplifier (OPA) of white-light continuum (Lightconversion ORPHEUS-F), pumped by a femtosecond Yb-based laser (Lightconversion PHAROS-SP-HP), generating pulses at 10 to 100 kHz repetition rate and $\lambda=1030$ nm, with Gaussian beam shape ($M^2$=1.3). The pulse shape was measured by using an autocorrelator (APE PulseCheck type 2), resulting in a sech temporal shape with pulse width ranging from 40 to 90 fs for the OPA, and between 0.14 to 7.9 ps for the main source. The OPA produces pulses at wavelengths between 650 and 940 nm (signal) and above 1160 nm, up to a few microns (idler). As shown in Fig.\ref{set-up}, the laser beam was focused by a 50 mm lens into the MMFs, with an input diameter at $1/e^2$ of peak intensity of approximately 30 $\mu$m at 1030 and 1500 nm and of 43.7 $\mu$m for the signal beam. Both standard 50/125 SIFs and 50/125 GIFs of variable length have been employed. The GIF core radius, core refractive index along the axis and relative core-cladding index differences at $\lambda = 1030$ nm are $r_c$=25 $\mu$m, $n_0$=1.472 and $\Delta$=0.0103, respectively.
The input tip of the fiber was imaged by a digital microscope (Dinolite-AM3113T), and light scattered out of the fiber was collected by means of a convex lens into a miniature fiber optics VIS-IR spectrometer (Ocean Optics USB2000+), working between 170 and 1100 nm. 
At the fiber output, a micro-lens images the out-coming near-field, and project it on a VIS-IR camera (Gentec Beamage-4M-IR). Besides monitoring the output transverse intensity distribution, the CCD also helps to optimize the input coupling. By a cascade of flipping mirror, the beam is then focused into an optical spectrum analyzer (OSA) (Yokogawa AQ6370D) and a real-time multiple octave spectrum analyzer (Fastlite Mozza) with wavelength ranges of 600-1700 nm and 1000-5000 nm respectively. Finally, the input and output average power was measured by a thermopile power meter (GENTEC XLP12-3S-VP-INT-D0). To check the presence of fiber damages induced by the propagation of too intense beams, we used a confocal optical micrscope (Zeiss Axio Scope.A1). 

\subsection{Optical nonlinear losses}
\begin{figure}[!h]
\centering\includegraphics[width=7.0cm]{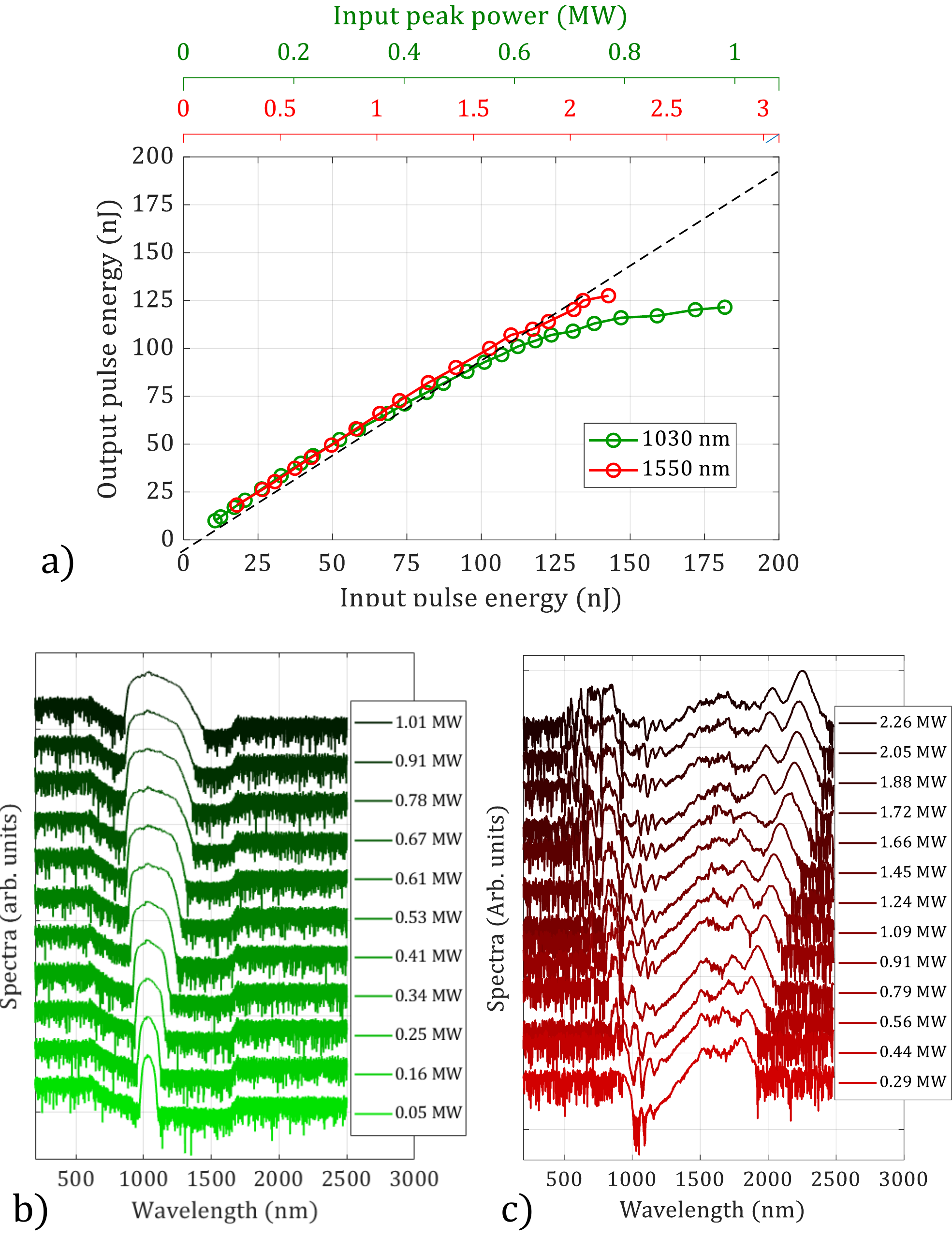}
\caption{a) Dependence of output pulse energy vs. the input energy, at $\lambda =$ 1030 nm (normal dispersion) or $\lambda =$ 1550 nm (anomalous dispersion) for 1 m of GIF. The dashed line represents the 100\% normalized transmission curve. 
b,c) Output spectra evolution for different input peak powers at 1030 (b) or 1550 nm (c), respectively. The corresponding input pulse duration is 174 fs and 61 fs, respectively. The visible range of the output spectra was collected by means of the same spectrometer that was used for characterizing PL.
}
\label{dispersion}
\end{figure}

Optical NLs consist of the transmission decrease when high-power pulses are injected into the fiber. This is shown in Fig.\ref{dispersion}a, where we report the input energy ($E_{in}$) dependence of the output pulse energy ($E_{out}$), for a 1-m long GIF. 
We investigated femtosecond pulse propagation in both the normal and in the anomalous dispersion regime, by using the pump at 1030 nm and the idler at 1550 nm, respectively. As Fig.\ref{dispersion}a shows, nearly the same behavior was found in both regimes.
Whenever $E_{in}< 80 nJ$, these measurements show that $E_{in}$ and $E_{out}$ are linearly proportional. This regime is characterized by the presence of weak linear losses, that are of the order of a few dB/km at 1030 nm \cite{Agrawal01}. Therefore, for short fibers (such as is the case in our experiments) losses due to material linear absorption can be neglected. For simplicity, we do not display injection losses due to coupling misalignment, so that 100\% transmission characterises the linear regime.
Fig.\ref{dispersion}a shows that, when $E_{in}> 80 nJ$, the output energy no longer scales linearly with input energy, resulting in a transmission drop, the hallmark of MPA-induced NLs.

In Fig.\ref{dispersion}b,c, we report the output spectra corresponding to the data of Fig.\ref{dispersion}a. As we can see, the two dispersion regimes lead to strongly different behaviors. At $\lambda = 1030$ nm, the output spectra broaden, but keep peaking at the source wavelength (see Fig.\ref{dispersion}b). Conversely, Fig.\ref{dispersion}c shows that at $\lambda = 1550 $ nm, a more complex dynamics occurs, leading to a significant enhancement of the nonlinear spectral broadening when compared with the case of Fig.\ref{dispersion}b. As a result of fission of the initial multisoliton pulse, Raman solitons are generated, which are strongly affected by MPA~\cite{zitelli2020high}. For wavelengths shorter than 1550 nm, anti-Stokes sidebands arise from spatiotemporal multimode soliton oscillations, and almost cover the entire visible range.

By comparing Fig.\ref{dispersion}a,b and c, we found that, in both dispersion regimes, MPA properties are remarkably similar, in spite of the profound differences in spectral and temporal dynamics. This permits us to justify the derivation, in Section 3, of a simple model for describing the effects of MPA on beam propagation, which only takes into account spatial effects.

\subsection{Fiber breakdown}
\begin{figure}[!h]
\centering\includegraphics[width=8.4cm]{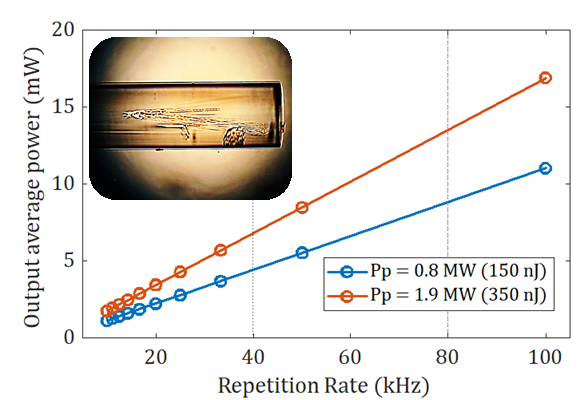}
\caption{
Average output peak power vs. the laser repetition rate, for 0.8 MW (linear loss regime) or 1.9 MW (NL regime) of input power. The laser wavelength and pulse duration were set to 1030 nm and 174 fs, respectively. The inset shows a microscope image of the input tip of a 50/125 GIF, after a laser beam with power right above the breakdown threshold was injected for a few minute.
}
\label{thermal}
\end{figure}

When operating at pulses intensities of the same order of magnitude of the laser-induced breakdown threshold, one naturally wonders whether irreversible processes may occur in the fiber. We verified that our experiments can be repeated several times, without observing any alteration of the fiber properties. However, in order to investigate the possible role played by thermal effects, we measured the fiber transmission when varying the laser repetition rate. We checked that the fiber output power linearly scales with the laser repetition rate, at peak powers in both the linear loss and in the NL regime (Fig.\ref{thermal}). This proves that thermal effects are negligible in our experiments. We also monitored the output transmission during the whole experiment, without noticing appreciable variations. 
On the other hand, we repeated the experiments with longer pulses, that are above 1 ps in temporal duration. Differently from the femtosecond regime, at the same values of peak power, we observed that the fiber transmission slowly but significantly drops in time. In particular, we monitored the transmission for several minutes, observing its progressive reduction (not shown), which is a sign of fiber breakdown. The input tip of the fiber was imaged by the optical confocal microscope working in cross-polarizer configuration in order to maximize the scattering from damages (see the inset of Fig.\ref{thermal}). As it can be seen, irreversible modifications of the sample were formed close to the input facet of the fiber, in proximity of the self-focusing point. 

\subsection{Role of spatial self-imaging}
\begin{figure}[ht!]
\centering\includegraphics[width=8.4cm]{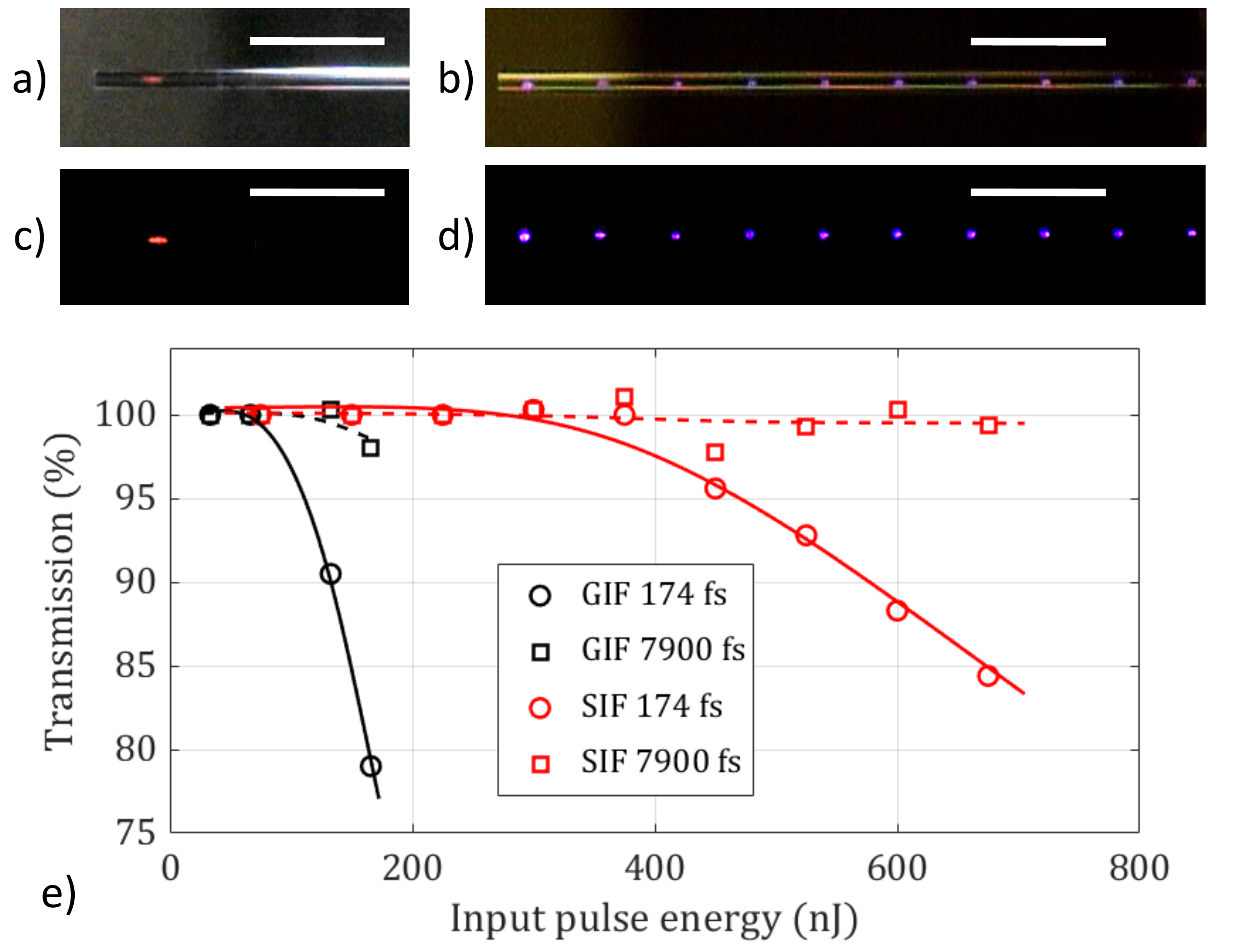}
\caption{a,b) Microscope images of the SIF (a) and the GIF (b) when the defects PL is excited by MPA of a 2 MW input peak power laser beam. c,d) Same as a,b), with room light switched off. The white bar is 1 mm long. e) Comparison between the two MMF transmissions, vs. input pulse energy, for a pulse duration of 174 fs (circle markers, solid lines) or 7.9 ps (square markers, dashed lines).}
\label{stepGIF}
\end{figure}

The results shown so far exclude the role of dispersion and thermal effects as the physical mechanism responsible for the femtosecond NL. Here, we show that the spatial evolution of the beam inside the fiber is the key element for activating the nonlinear optical attenuation. For doing so, we compare the nonlinear transmission properties of MMFs with the same core/cladding size, with (GIF) or without (SIF) SSI.
As a result of MPA, one observes the up-conversion PL of silica defects, which, thanks to its typical violet color, helps tracking the spatial dynamics of the beam inside the MMFs. This permits to visualize the different beam evolutions which take place inside the SIF and the GIF. 
Optical beams propagating in a SIF experience a single self-focusing point at the very beginning of the fiber (see the digital microscope image of PL in Fig.\ref{stepGIF}a,c). The resulting high intensity leads to MPA, so that most of the beam energy is lost over the first few millimeters of propagation. Conversely, in a GIF the beam diameter periodically oscillates in space, owing to SSI \cite{krupa2019multimode}. The associated spatial beam breathing produces several minima for the beam diameter, where MPA may lead to NL. As a result, as shown in Fig.\ref{stepGIF}b,d, PL appears as an array of equally spaced emitting points \cite{hansson2020nonlinear}.
The different beam dynamics for the two types of MMF have a direct consequence on power transmission measurements. In Fig.~\ref{stepGIF}e, we report the optical transmission as a function of the input pulse energy, for input pulse durations of either 174 fs or 7.9 ps, with $\lambda = 1030$ nm. As it can be seen, for both MMFs the transmission curves remain flat when picosecond pulses are used. On the other hand, using femtosecond input pulses lead to observing strong NL. Moreover, the threshold for a transmission drop is quite different for the two MMFs. As shown in Fig.\ref{dispersion}e, a drop occurs above 80 nJ of input pulse energy in a GIF, whereas a fourfold threshold enhancement is observed in a SIF. For this reason, experiments carried out with a GIF will be discussed in the following subsection.


\subsection{Spectral analysis}
\begin{figure}[ht!]
\centering
\includegraphics[width=8.4cm]{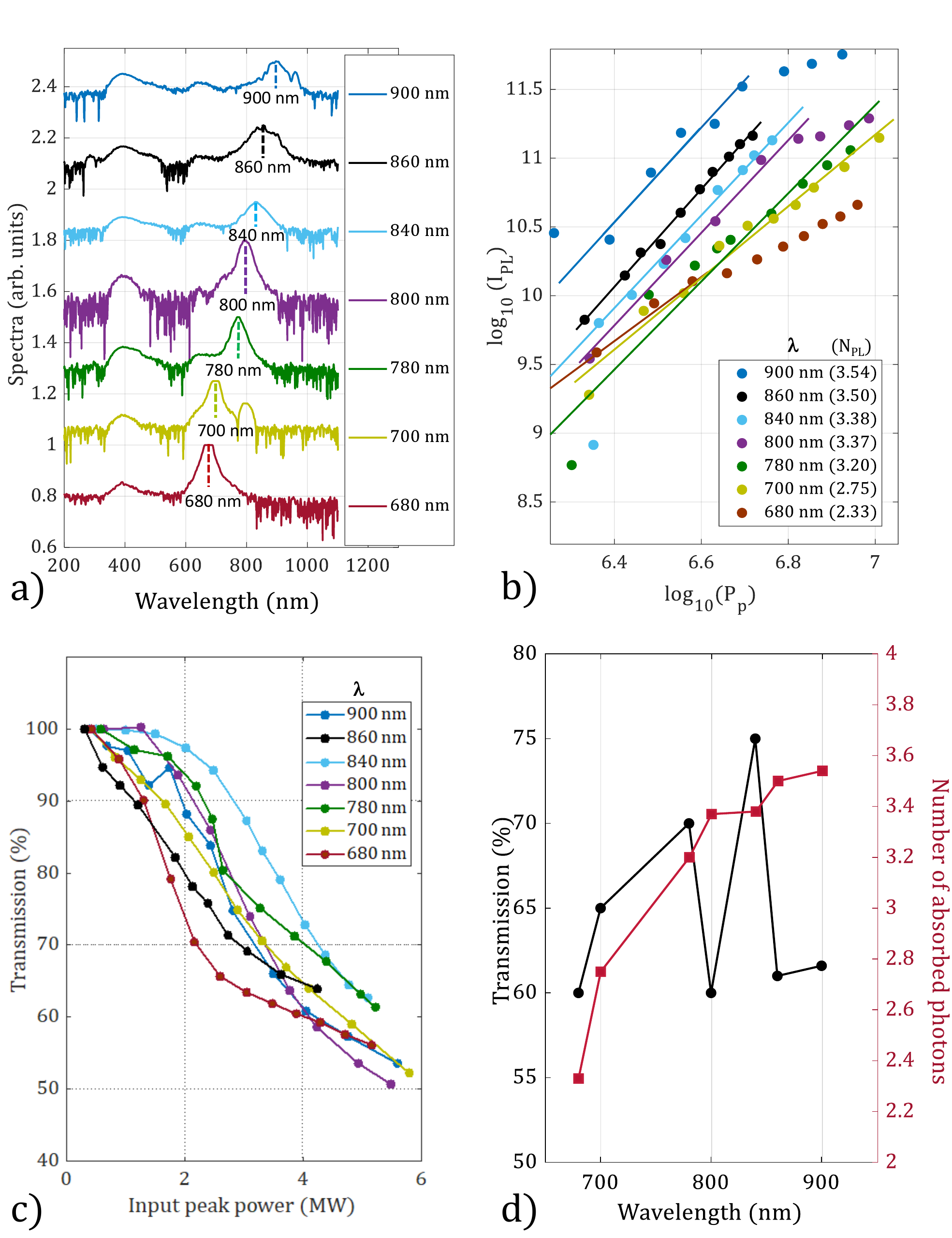}

\caption{(a) Side-scattered spectra for different source wavelengths, at $P_p$ = 2.5 MW of input peak power.
(b) Log-log plot of the Ge-ODC PL intensity vs. $P_p$. In the legend, the number in brackets denotes the calculated $N_{PL}$. 
(c) Fiber NL for different source wavelengths.
(d) Comparison between the fiber transmission at $P_p$ = 4 MW and $N_{PL}$, vs. source wavelength.
}
\label{spect}
\end{figure}

With the aim of investigating the physical origin of MPA, we studied how both PL and fiber transmission vary with the source wavelength. As depicted in the set-up of Fig.\ref{set-up}, we collected the PL signal into a VIS-IR spectrometer. In Fig.\ref{spect}a we show measured spectra for a laser peak power of 2.5 MW, and wavelength varying between 680 and 900 nm. Besides the spectral broadening of the pump, observed spectra display the characteristic NBOHC and the Ge-ODC PL peaks, occurring at 650 and 400 nm, respectively. Following the same method of ref.\cite{ManginiFerraro2020PRApp}, we suppose a power-law linking the Ge-ODC PL intensity ($I_{PL}$) and the input peak power ($P_p$): $I_{PL} \propto P_p^{N_{PL}}$, where $N_{PL}$ is the average number of photons involved in the MPA process exciting the defects PL. Thus $N_{PL}$ can be obtained by evaluating the slope of the linear dependence from $P_p$ of $I_{PL}$ (calculated as the integral of the corresponding peak in the spectrum), when displayed in a log-log plot. As reported in Fig.\ref{spect}b, the value of $N_{PL}$ varies with the source wavelength. Specifically, about 3 photons are simultaneously absorbed at $\lambda =$ 750 nm, since the Ge-ODC absorption band has a maximum at about 250 nm \cite{girard2019overview}. We do not report here the corresponding analysis for NBOHC, since its contribution to PL so small, that the resulting linear fit is too noisy to provide reliable estimation of $N_{PL}$. Furthermore, the NBHOC PL signal partially overlaps with the source at small wavelengths, as it can be seen in Fig.\ref{spect}a.

Next, we compared the analysis of the wavelength dependence of side-scattered spectra with that of fiber transmission measurements. As shown by Fig.\ref{spect}c, NL was observed at all source wavelengths. Nonetheless, differently from the PL related curves of Fig.\ref{spect}b, the transmission curves are not sorted with respect to the input wavelength. This indicates that Ge-ODC and NBOHC absorption is not the only mechanism which is responsible for NLs. For a clearer comparison, in Fig.\ref{spect}d we simultaneously plot the wavelength dependence of both $N_{PL}$ and of the transmission at $P_p = 4 MW$.  As can be seen, while $N_{PL}$ monotonically increases with the source wavelength, the transmission curve exhibits a complex behavior. Specifically, two transmission maxima were obtained around 750 and 840 nm, respectively. The first maximum corresponds to three times the Ge-OCD and NBOHC absorption band peaks. Whereas the second maximum can be ascribed to other types of defects that do not contribute to PL, such as the Ge(1) and Ge(2) centers, whose associated oscillation strengths are comparable to those of luminescent defects \cite{girard2019overview}.

\subsection{Macroscopic spatial evolution}
\begin{figure}[ht!]
  \centering
\includegraphics[width=8.4cm]{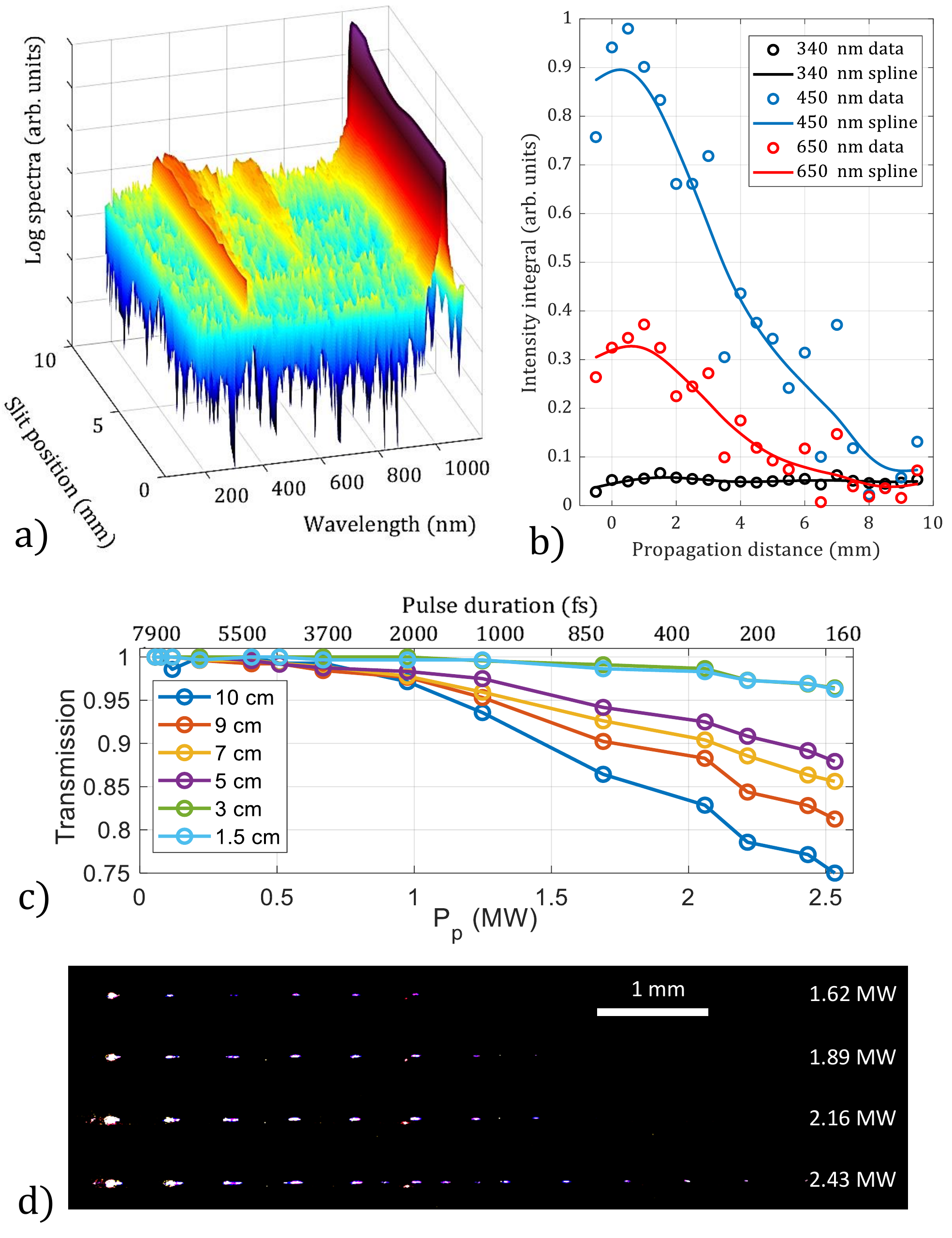}
\caption{a) Side-scattering spectrum, obtained when varying the slit position. b) Integral of the spectral peaks in a). Solid lines are a guide for the eye. c) Cutback experiment from 10 cm to 1.5 cm of fiber length. The $P_p$ value was varied by changing the input pulse duration between 7.9 ps and 174 fs, while keeping the pulse energy unchanged. d) PL intensity variation along with $P_p$. Images from the top to the bottom correspond to $P_p =$ 1.62, 1.89, 2.16 and 2.43 MW respectively. 
}
\label{cutback}
\end{figure}

In order to track the spatial evolution of PL along the beam propagation, we placed the spectrometer and the lens on a translation stage, which allows for shifting them solidly. In Fig.\ref{cutback}a we show the longitudinal evolution of the side-scattered light spectrum, over the first centimeter of a GIF, for 2.5 MW of input peak power. Besides the pump at 1030 nm, three additional main peaks appear: the third harmonic (TH) at 343 nm, and two PL peaks at 400 and 650 nm, respectively. 
The resulting collected spectra were post-processed, in order to determine the evolution of the spectral energy associated with each peak. Fig.~\ref{cutback}.b shows that the energy of PL peaks significantly drops over less than 1 cm of fiber. On the other hand, our measurements (not shown here) reveal that the TH (and the pump) energy is only damped over several centimeters (or several tens of centimeters). 
The observed decrease of the PL is associated with a drop of the peak intensity of the beam. However, the drop of PL intensity does not necessarily prove a loss of the beam energy. For example, an intensity decrease may be due to a temporal broadening of the pulses, as well as to an increase of the effective area of the beam. However, the observed rapid damping of light side-scattered by the pump indicates that, even if present, these contributions are negligible (at least for fiber lengths of the order of 1 m, as in Fig.\ref{dispersion}a). 

In order to verify that indeed MPA processes are responsible for the observed PL signal drop of Fig.~\ref{cutback}.b, we performed a cut-back experiment, by progressively reducing the GIF length from 10 down to 1.5 cm, while keeping the input coupling conditions unchanged. The measured transmission as a function of $P_{p}$ is reported in Fig.\ref{cutback}c. $P_p$ was varied by changing the laser compressor ratio. Namely, thanks to a feedback control of the laser system, the energy of each pulse was kept a constant, while the pulse duration was varied between 140 fs and 7.9 ps. The advantage of this method is that optical elements that may introduce small misalignments, such as variable attenuators, are not needed. As previously observed with reference to Fig.\ref{dispersion}a, two distinct power transmission regimes are revealed. For powers below the threshold value of $P_p=0.5$ MW, the fiber transmission remains nearly constant for all fiber lengths. Whereas at higher powers, the transmission starts decreasing. It is interesting to point out that such a power threshold coincides with the value that leads to the appearance of PL (see Fig.\ref{cutback}d). This observation suggests that MPA leadingto PL is one of the main mechanisms which are responsible for NLs. However, other nonlinear effects, such as the aforementioned non luminescent defects excitation, may give a contribution to NLs. In the next section, we propose a model based on an effective N-photon absorption, which is able to quantitatively reproduce the experimental transmission drop.




\section{Analytical model}
\begin{figure}[ht!]
  \centering
  \includegraphics[width=7.5cm]{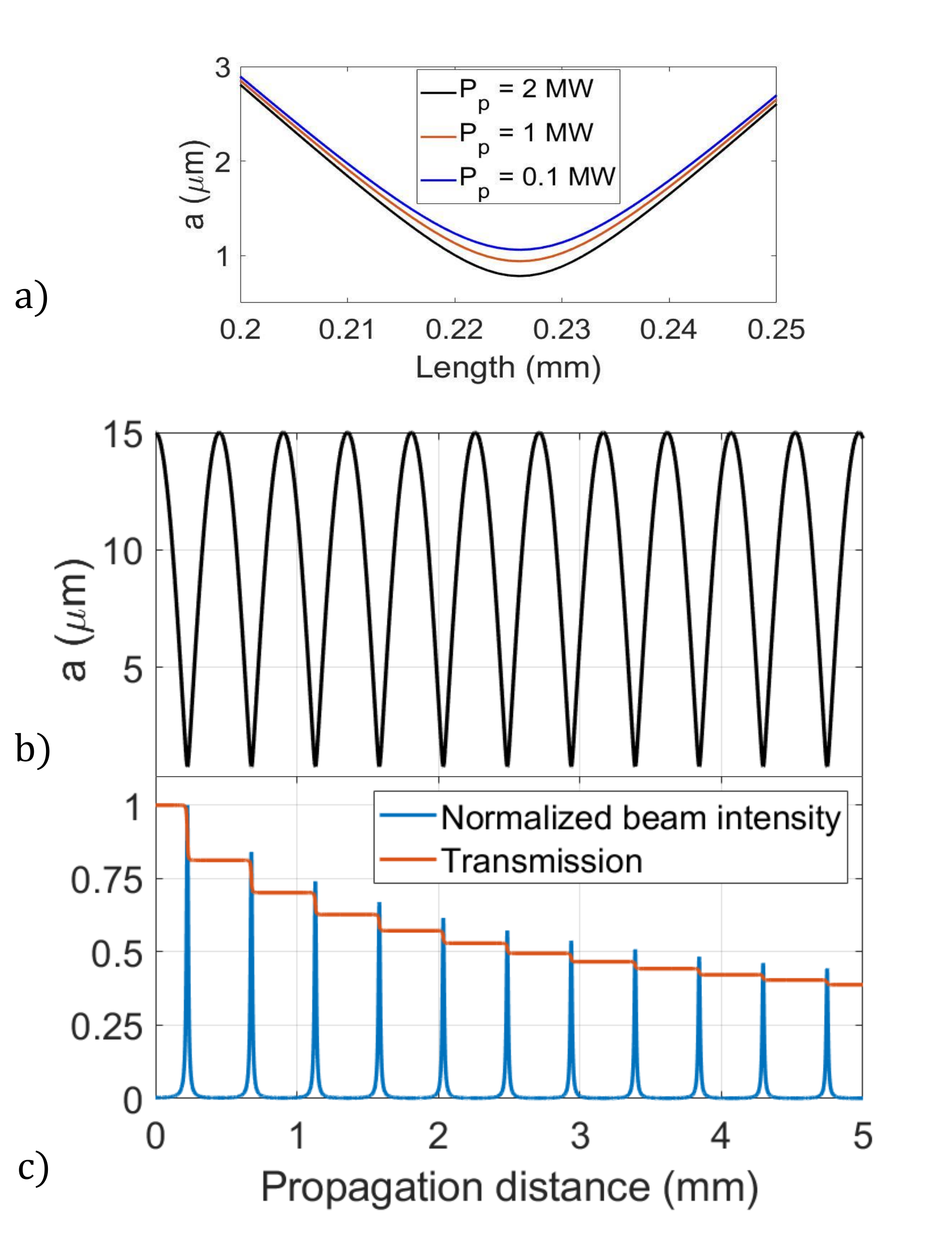}
\caption{a) Detail of the beam size minimum, for different values of $P_p$.
b) Beam size oscillation along the propagation distance, for $P_p = 2$ MW. c) Evolution of beam intensity (normalized to its maximum value) and transmission along the first 5 mm of GIF, as obtained from the N-photon absorption model in Eq.(\ref{eq-fit}) with $N = 3$ and $\alpha_3 = 10^{-31} m^3/W^2$: }
\label{model}
\end{figure}

Let us consider the beam spatial dynamics under the slowly-varying envelope approximation, treating time as a parameter, analogously to the variational approach proposed by Karlsson in 1992 \cite{karlsson1992dynamics}. In order to take into account the experimental NL, we added to the index grading and the Kerr nonlinearity, an effective $N$ photon absorption term into the GIF permittivity:
\begin{equation}
    \varepsilon \simeq n_0^2 + 2 n_0 n_2|E|^2-gn_0^2r^2+i n_0\frac{\alpha_N}{ k_0}|E|^{2N-2}.
\end{equation}
Here, $n_2$ is the nonlinear refractive index, $E$ is the envelope of the propagating electric field in the fiber (measured in $\sqrt{W/m}$), $g$ is the index grading parameter defined as $g=2n_0\Delta/r_c^2$, $r$ is the independent radial coordinate, $k_0=2\pi n_0/\lambda$ is the propagation constant in the core and $\alpha_N$ is the N photon absorption coefficient. We highlight that $N$ represents the average number of photons involved in the nonlinear absorption process, that sums all possible contributions to energy losses, including side-scattering.
As in \cite{karlsson1992dynamics}, we impose a Gaussian shape of the beam, so that the electric field can be written as $E(z,r)=A(z)e^{-\sigma(z)r^2}$. However, in a nonconservative system the Lagrangian equations of the variation approach do not hold, and the beam power defined as 
\begin{equation}
P=2 \pi \int|E|^2 r dr
\end{equation}
is no longer an integral of motion. Here, we consider an alternative method to the Lagrangian equations: we impose that only the $z$ dependent amplitude $A$ is affected by absorption, so that $\sigma$ is independent of $\alpha_N$. In this way, we can recover the evolution of $\sigma$ from the lossless problem. With this consideration, the beam power evolution equation reads as (see Supplementary materials for the step-by-step derivation)
\begin{equation}
    \frac{dP}{dz}=-\frac{\alpha_N}{2N-1}\frac{P^N}{(\pi a^2)^{N-1}},
    \label{eq-fit}
\end{equation}
where $a(z)$ is the beam $1/e^2$ radius that oscillates along $z$ due to the SSI, according to the formula:
\begin{equation}
    a(z)=a_0\sqrt{\cos^2(\sqrt{g}z)+C\sin^2(\sqrt{g}z)}.
\end{equation}
The parameter $C$ contains the dependence of the beam size upon input peak power, and it is defined as
\begin{equation}
    C=\frac{1-p}{k^2a_0^4g},
\end{equation}
with $p={n_2k^2P_p}/{2\pi n_0}$ and $P_p = \pi A_0^2 a_0^2$. The fiber breakdown condition is given by imposing $C=0$, which corresponds to the peak power threshold $P_p^{thr.} = 4.26$ MW (using the experimental parameters at $\lambda= $ 1030 nm and $n_2=2.7\cdot10^{-20} m^2/W$). While the SSI period is power independent, the beam size oscillation amplitude depends on $P_p$ (as depicted in Fig.\ref{model}a)\cite{hansson2020nonlinear}. In Fig.\ref{model}b, we report the oscillating behavior of the beam size along the propagation distance, for the power $P_p = 2$ MW, right below the breakdown threshold. 

In an MPA process, the energy loss strongly depends on the beam waist through its intensity. At points of maximum intensity, the MPA contribution becomes most relevant. Even though the absorption coefficient is small, the intensity locally reaches such high values that the NL term in equation (\ref{eq-fit}) is no longer negligible. This results in a series of step-wise drops of the transmission curve, at points of SSI-induced beam focusing, as shown in Fig.\ref{model}c. 

\subsection{Cutback experiment fitting}
As a last result, we fitted the experimental data of the cutback measurements in Fig.\ref{cutback}c with our analytical model (see Supplementary materials for details). As shown in Fig.\ref{fit}a, we obtained a good quantitative agreement: the fitting function interpolates well experimental data for all values of $P_p$ and $z$. The extrapolated parameters are $N =$ 3.008 and $\alpha_N = 2.415 \cdot 10^{-33}$ (in SI units). As already foreseen in the comparison between the transmission curves and $N_{PL}$, a value of $N \simeq 3$ indicates that MPA which excites PL is not the sole responsible for the observed NL. Many other factors contribute to the value of $N$, such as: Ge(1) and Ge(2) defects absorption,
and generation of the pump TH, which is not guided (and therefore it is laterally scattered from the fiber, as shown in Fig.\ref{spect}a,b). As a general remark, our model is rather simple, since it considers a monochromatic wave, in spite of the broad spectra of Fig.\ref{dispersion}b. Furthermore, we are not considering effects such as multiphoton ionization \cite{stuart1996nanosecond}, color-center generation \cite{efimov1998color}, tunneling photoionization \cite{wu2005femtosecond}, and plasma formation \cite{cho2002situ}, which have been shown to play a role when MW peak power femtosecond lasers are employed. Models of femtosecond laser absoption in silica which take into account filamentary propagation, ablation and refractive index micro-modification can be found in literature \cite{sudrie2002femtosecond,dostovalov2015quantitative,stuart1995laser}. Nevertheless, we believe that our model achieves its purpose, as it finds a remarkably good agreement with the experimental data. Specifically, we could reveal that an effective 3-photon absorption term is useful for describing the beam dynamics in a variational-like approach, when working close to the fiber breakdown threshold.

\begin{figure}[ht!]
  \centering
  \includegraphics[width=7.5cm]{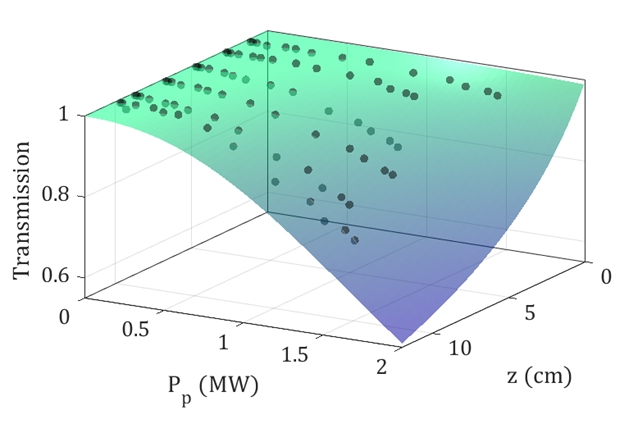}
\caption{Fit of the cutback experimental data in Fig.\ref{cutback}d with the model in eq.(\ref{eq-fit}). The fit parameters are $N=3.008$ and $\alpha_N = 2.415 \cdot 10^{-33}$.}
\label{fit}
\end{figure}

\section{Conclusion}
We experimentally characterized nonlinear optical losses in MMFs, when operating slightly below the breakdown power threshold. In this regime, MPA mechanisms play a significant role, and may strongly affect the dynamics of propagating beam. For example, in the anomalous dispersion regime it has been shown that MPA clamps the beam output energy, thus suppressing the Raman soliton self-frequency shift \cite{zitelli2020high}. In this work, based on the observed close similarity between nonlinear power transmission properties of MMFs in both the normal and in the anomalous dispersion regime, we carried out a detailed investigation of the spatial beam dynamics. Specifically, we revealed that SSI has a key role in enhancing NL. Therefore, MPA effects are remarkably higher in GIFs than in SIFs. By comparing side-scattering and transmission measurements, we could infer that the observed NL cannot be entirely ascribed to the MPA which is associated with defects PL. Although all of the different sources of NL could not be directly identified, we demonstrated that the observed nonlinear transmission drop could be quantitatively well reproduced by introducing an effective N-photon absorption term in the propagation equation. Namely, we found that $N= 3$ at $\lambda=$ 1030 nm provides a good agreement with experiments carried out at several input peak powers and fiber lengths. Our results highlight the intrinsic limitation of VIS-IR light propagation in MMFs due to MPA, which becomes highly relevant when propagating femtosecond pulses of MW peak powers, i.e., close to, but still below, the fiber breakdown threshold. In this sense, our results will be of significant interest for different applications based on MMFs, such as micro-machining, medical imaging, beam delivery, and fiber lasers for spatiotemporal mode-locking. In all of these emerging technologies, MPA represents a major obstacle for their power up-scaling.

\begin{backmatter}

\bmsection{Acknowledgments}
We acknowledge support from the European Research Council (ERC) under the European Union’s Horizon 2020 research and innovation program (grant No. 740355 and grant No. 874596), and the Italian Ministry of University and Research (R18SPB8227).
The authors declare no conflicts of interest.

\bmsection{Disclosures}
The authors declare that they have no competing financial interests.

\bmsection{Supplemental document}
See Supplement 1 for supporting content. 
\end{backmatter}

\bibliography{bibliog}






\newpage

{\Huge Supplementary Materials}
\section*{Theoretical Model}

In this section we show the step-by-step derivation of Eq.(3) in the main text. In our modelling of the optical properties of a GIF, we take into account Kerr effect, $N$ photon absorption and the index grading, so that the permittivity $\varepsilon$ reads
\begin{equation}
    \varepsilon=\Bigg(n_0+n_2|E|^2+i\frac{\alpha_N}{2 k_0}|E|^{2N-2}-g\frac{r^2}{2}n_0\Bigg)^2 \qquad\qquad g = 2 n_0\frac{\Delta}{r_c^2}
\end{equation}
Its approximated form is reported in the main text as
\begin{equation}
    \varepsilon \simeq n_0^2 + 2 n_0 n_2|E|^2+i n_0\frac{\alpha_N}{ k_0}|E|^{2N-2}-gn_0^2r^2.
\end{equation}
In order to study the beam dynamics we write the Helmholtz equation in polar coordinates
\begin{equation}
\frac{1}{r}\frac{\partial}{\partial r}\Bigg ( r\frac{\partial E}{\partial r}\Bigg ) + 2 i k \frac{\partial E}{\partial z}-k^2r^2gE+2\frac{n_2}{n_0}k^2|E|^2E+ik\alpha_N|E|^{2N-2}E=0,
    \label{Helmholtz}
\end{equation}
where $k=k_0n_0$ is the beam wave vector in the fiber. 
Analogously to the variational approach \cite{karlsson1992dynamics}, we force the beam to own a spatial Gaussian profile at all the propagation distances $z$:
\begin{equation}
    E(z,r)=A(z)e^{-\sigma r^2}
    \label{gauss}
\end{equation}
Substituting Eq.(\ref{gauss}) in Eq.(\ref{Helmholtz}) the wave equation becomes:
\begin{equation}
\begin{split}
    4\sigma(\sigma r^2 - 1)Ae^{-\sigma r^2}+2ik\Bigg(\frac{dA}{dz}-r^2A\frac{d\sigma}{dz}\Bigg)e^{-\sigma r^2}-k^2r^2gAe^{-\sigma r^2}+\\+2\frac{n_2}{n_0}k^2|A|^2Ae^{-3\sigma r^2}+ik\alpha_N|A|^{2N-2}Ae^{-(2N-1)\sigma r^2}=0
\end{split}
\end{equation}
Now, the variable $r$ is explicit and it can therefore be easily integrated, giving
\begin{equation}
    2i\Bigg(\frac{1}{2\sigma}\frac{dA}{dz}-\frac{A}{2\sigma^2}\frac{d\sigma}{dz}\Bigg)-kg\frac{A}{2\sigma^2}+2\frac{n_2}{n_0}k|A|^2A\frac{1}{6\sigma}+i\alpha_N|A|^{2N-2}A\frac{1}{2(2N-1)\sigma}=0.
\end{equation}
Multiplying by $\sigma^2$ we get
\begin{equation}
    2i\Bigg( \sigma \frac{dA}{dz}-A\frac{d\sigma}{dz}\Bigg)-kgA+\frac{2}{3}\frac{n_2}{n_0}k\sigma|A|^2A+i\alpha_N\frac{\sigma}{2N-1}|A|^{2N-2}A=0
\end{equation}
At this point, we explicit real and imaginary parts of $\sigma$ and $A$:
\begin{equation}
    \sigma = \Re{\sigma} + i \Im{\sigma}; \qquad\qquad A = |A|e^{i\phi}\qquad\qquad \frac{dA}{dz}=e^{i\phi}\Bigg(\frac{d|A|}{dz}+i|A|\frac{d\phi}{dz}\Bigg)
\end{equation}
so that the wave equation becomes
\begin{equation}
\begin{split}
    2i\Bigg[(\Re{\sigma}+i\Im{\sigma})\Bigg(\frac{d|A|}{dz}+i|A|\frac{d\phi}{dz}\Bigg)-|A|\Bigg(\frac{d\Re{\sigma}}{dz}+i\frac{d\Im{\sigma}}{dz}\Bigg)\Bigg]-kg|A|+\\+\frac{2}{3}\frac{n_2}{n_0}k(\Re{\sigma}+i\Im{\sigma})|A|^3+i\frac{\alpha_N}{2N-1}(\Re{\sigma}+i\Im{\sigma})|A|^{2N-1}=0
    \end{split}
\end{equation}
We can now separate real and imaginary part, obtaining two coupled equations\\
Re:
\begin{equation}
    -2\Im{\sigma}\frac{d|A|}{dz}-2\Re{\sigma}|A|\frac{d\phi}{dz}+2|A|\frac{d\Im{\sigma}}{dz}-kg|A|+\frac{2}{3}\frac{n_2}{n_0}\Re{\sigma}k|A|^3-\frac{\alpha_N}{2N-1}\Im{\sigma}|A|^{2N-1}=0
    \label{real}
\end{equation}
Im:
\begin{equation}
    2\Re{\sigma}\frac{d|A|}{dz}-2\Im{\sigma}|A|\frac{d\phi}{dz}-2|A|\frac{d\Re{\sigma}}{dz}+\frac{2}{3}\frac{n_2}{n_0}\Im{\sigma}k|A|^3+\frac{\alpha_N}{2N-1}\Re{\sigma}|A|^{2N-1}=0
    \label{imag}
\end{equation}
In order to get rid of $\phi$, we combine the previous two equation, calculating $\Im{\sigma}\cdot(\ref{real})-\Re{\sigma}\cdot(\ref{imag})$
\begin{equation}
    \frac{d|A|}{dz}=|A|\Bigg(\frac{d}{dz}\ln |\sigma| - kg\frac{\Im{\sigma}}{2|\sigma|^2}-\frac{\alpha_N}{2(2N-1)}|A|^{2N-2}\Bigg),
\end{equation}
that can be rewritten as follow
\begin{equation}
    \frac{d|A|}{dz}=|A|\Bigg ( f(z)-\frac{\alpha_N}{2(2N-1)}|A|^{2N-2}\Bigg),
    \label{sup-dA}
\end{equation}
where $f(z)$, that is unknown at this point, includes the dependence from $\sigma$. In order to find an expression for $f(z)$, we can recall the lossless case, in which the beam power defined as
\begin{equation}
P=2 \pi \int|E|^2 r dr
\end{equation}
is an integral of motion, i.e. $dP/dz = 0$. Dubbing $a$ the $1/e^2$ beam radius in the lossless problem, the condition $dP/dz = 0$ returns 
\begin{equation}
    f(z)=-\frac{1}{a}\frac{da}{dz},
    \label{sup-f}
\end{equation}
because
\begin{equation}
        P(z)=\pi |A|^2a^2.
        \label{power}
\end{equation}
Deriving Eq.(\ref{power}) respect to $z$ and recalling Eq.(\ref{sup-dA}) and Eq.(\ref{sup-f}), one eventually gets:
\begin{equation}
    \frac{dP}{dz}=-\frac{\alpha_N}{2N-1}\frac{P^N}{(\pi a^2)^{N-1}}.
    \label{first-result}
\end{equation}
that is Eq.(3) in the main text.

\subsection*{Cutback experiment fitting}
Eq.(\ref{first-result}) can be solved with the initial condition $P(z=0)=P_p$ by separating the variables $z$ and $P$
\begin{equation}
P(z) = \frac{P_p}{\Bigg[1+\frac{N-1}{2N-1}\alpha_N P_p^{N-1}\int_{0}^z\frac{dz'}{[\pi a^2(z')]^{N-1}}\Bigg]^{\frac{1}{N-1}}}.
\label{solution}
\end{equation}
The Matlab routines that we used for the fitting of the experimental data is unable to compute Eq.(\ref{solution}) as the length of the variable $z$ of the fit must be chosen dependently on a fit parameter (N). This is needed from the routine in order to calculate the integral. Therefore, we cannot use Eq.(\ref{solution}) to fit the experimental data. However, a reasonable approximation can be done to get rid of the integral. In fact, since $a$ oscillates very fast with respect to the value of the experimental lengths, we can approximate the integral in (\ref{solution}) as the ratio between $z$ and the oscillation period $\pi/\sqrt{g}$ multiplied by the integral over one period.
\begin{equation}
    \int_0^z \frac{dz'}{[\pi a^2(z')]^{N-1}} \simeq \frac{\sqrt{g} z}{\pi}\int_0^{\pi/\sqrt{g}} \frac{dz'}{[\pi a^2(z')]^{N-1}}.
    \label{approx}
\end{equation}
With this approximation, the solution (\ref{solution}) can be written as
\begin{equation}
    P(z) = \frac{P_p}{\Bigg[1+\frac{N-1}{2N-1}I_{N-1}\alpha_N\Big(\frac{Pp}{\pi a_0^2} \Big)^{N-1} z \Bigg]^{\frac{1}{N-1}}},
\end{equation}
where
\begin{equation}
    I_N=\frac{1}{\pi}\int_0^\pi \frac{dy}{[\cos^2 y + C \sin^2 y]^N},
\end{equation}
and the Matlab fitting routines run. One may notice that with the approximation (\ref{approx}), we cannot reproduce the step-wise trend of the transmission as in Fig.7c of the main text. Indeed, the oscillating behavior is integrated over a fixed length. In any case, it worth noting that getting experimental data dense enough to fit such an oscillation would require a micron precision cleaving technique, which would be practically impossible with our tools.

\end{document}